\documentclass{llncs}
\usepackage[latin1]{inputenc}
\usepackage[english]{babel}
\usepackage{graphicx}
\usepackage{subfigure}

\begin{document}

\title{Cellular Automata under the influence of noise}
\titlerunning{Noise in CA}  
%
\author{Luís Correia\inst{1,2} \and Thomas Wehrle\inst{3}}
\authorrunning{Correia and Wehrle}   
%
\tocauthor{Luís Correia (LabMAg---University of Lisbon),
Thomas Wehrle (University of Zurich)}

\institute{LabMAg - University of Lisbon,\\
  Campo Grande 1749-016, Lisboa, Portugal\\
\email{Luis.Correia@di.fc.ul.pt}
\and
AI Lab - University of Zurich,\\
Andreasstrasse 15, CH-8050, Zürich, Switzerland
\and
University of Zurich, Institute of Psychology,\\
Zürichbergstrasse 43, CH-8044, Zürich, Switzerland\\
\email{t.wehrle@psychologie.unizh.ch}}

\maketitle              

\begin{abstract}
  Noise in the local transition function is compared to fluctuations
  in the updating times of the cells.  Obtained results are shown to
  be quite different in both cases.  In this extended abstract we
  briefly explain the problem and present results obtained and comment
  them.
\end{abstract}
\section{Introduction}
In their original definition, Cellular Automata (CA) are synchronous
\cite{smi:76}.  This means that all cells are updated in parallel
simultaneously.  However it has been shown that this updating policy
induces artificial structure in the CA states
\cite{hog:80,ing_buv:84,ber_det:94,sch_roo:99}.  Moreover, natural
collective systems are not synchronous, in the sense that their
components do not react to a signal exactly at the same time
\cite{cor_al:02}.  Therefore, asynchronous CA have been studied and
defended as more realistic models of natural systems
\cite{hog:88,lc:06}.

By comparison to synchronous models, asynchronous ones may be regarded
as having noise \cite{kan:97,sch_roo:99}.  In fact, we may consider
that besides signal amplitude fluctuations, timing fluctuations are
also a form of noise \cite{lc:06b}.  In the latter case it takes the
form of delays or speedups in the updating time instead of changes in
the output of the local function.

It has been hypothesised that these two forms of noise may
qualitatively produce the same or a similar effect
\cite{sch_roo:99,rux_sar:98}.  In this paper we show that this is not
the case.  Qualitatively different results are produced by noise in
the local function and by noise in the updating moments.  We used 1D
binary CA with neighbourhood of radius 1.

The next section presents the models considered, the following one
shows results obtained and the paper ends with an analysis to the
results and ideas to develop further.

\section{Models}

In asynchronous CA, several updating policies are possible, producing
different results \cite{kan:97,sch_roo:99} and possessing more or less
realism.  According to a statistical analysis in \cite{sch_roo:99} the
most realistic asynchronous updating model is time driven, with
independent timings for each cell.  The waiting times of each timer
are exponentially distributed, with mean 1.

The step-driven method with uniform choice is shown to produce
approximate results for large CA grids.  It just randomly picks the
next cell to be updated from all the cells in the grid.  Therefore,
for a grid with $N$ cells, after $N$ updates, some cells may have been
updated more than once while others may not have been updated at all.
This method is simpler to compute and may also be called
\emph{unfair}.  Nevertheless, over time all cells will tend to equal
number of updates. It has been shown \cite{sch_roo:99} that for
updates in which the local function is independent of time, this model
is equivalent to the time driven above described.

Fairness in each set of $N$ updates may be forced originating the
asynchronous step-driven random new sweep method \cite{sch_roo:99}.
In this model the $N$ cells are randomly sorted and this gives the
update order for the next individual $N$ updates.  Therefore, in a
cycle of $N$ updates, all the cells are updated exactly once, although
in a different random order in each cycle.  For this reason we may
designate this updating by \emph{fair}.

The fair model, in spite of the artificiality of guaranteeing exactly
one update per each cycle of $N$, may be quite realistic for models of
natural systems.  We may consider that the result of $N$ updates is
just a snapshot that we may observe cyclically.  If each component of
a natural system has a latency time (minimum time between two
successive state changes) that is long compared to the state changing
time and the state changes roughly self-synchronise among cells, then
the fair model may be quite realistic.  It should be noticed that
self-synchronisation in natural systems is quite a frequent process
(see for instance \cite{cor_al:02}).

The synchronous model is the commonly used.  Each cell update is
computed for all cells but the new states are only loaded into the
cells after all updates performed.

For any of this updating models we may consider noise in the local
function. It the most general form, for these CA, it may introduce an
error in the output with two given probabilities, $p_{10}$ for an
output error of 1 to become a 0 and $p_{01}$ for the other way round.

\section{Results}

We tested the three models with and without errors in the local
function.  Results are not qualitatively different between the two
asynchronous updates (which confirms results obtained in
\cite{cor_al:02} without error in the local function).  Therefore, we
only present a comparison between the asynchronous unfair and the
synchronous updatings.

In introducing an error in the local function we opted for considering
only errors from an active to a quiescent state.  Otherwise, the error
would introduce a kind of seed from which active structures could be
built in the CA.  In modeling of natural systems this is an important
issue.  Simulated systems have a kind of background (the quiescent
state) over which the action takes place.

\begin{figure}[center]
  \centering
  \subfigure[sync]{
    \label{fig:110:a}
    \includegraphics[width=.21\textwidth]{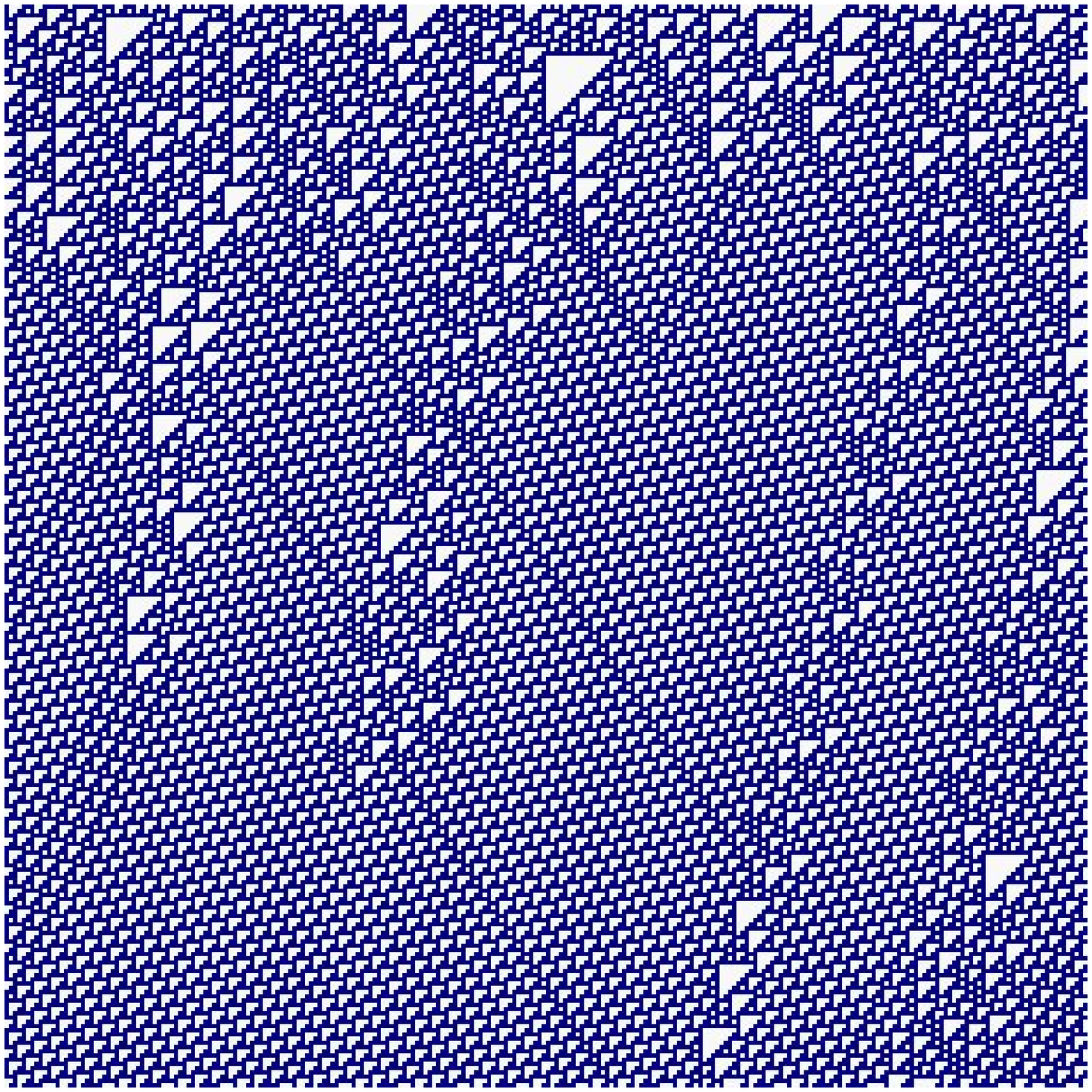}}
  \subfigure[sync err]{
    \label{fig:110:b}
    \includegraphics[width=.21\textwidth]{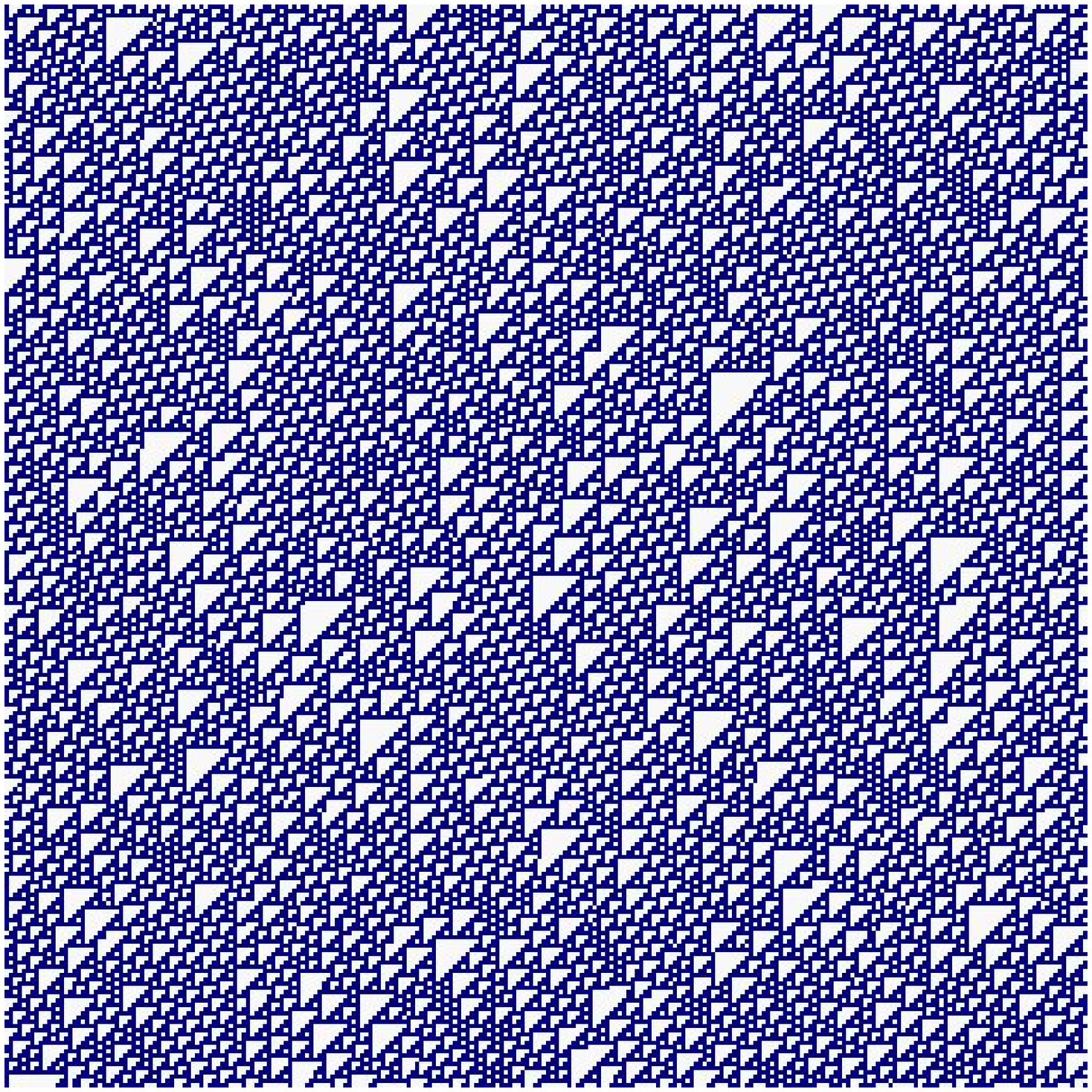}}
  \hspace{0.3cm}
  \subfigure[unfair]{
    \label{fig:110:c}
    \includegraphics[width=.21\textwidth]{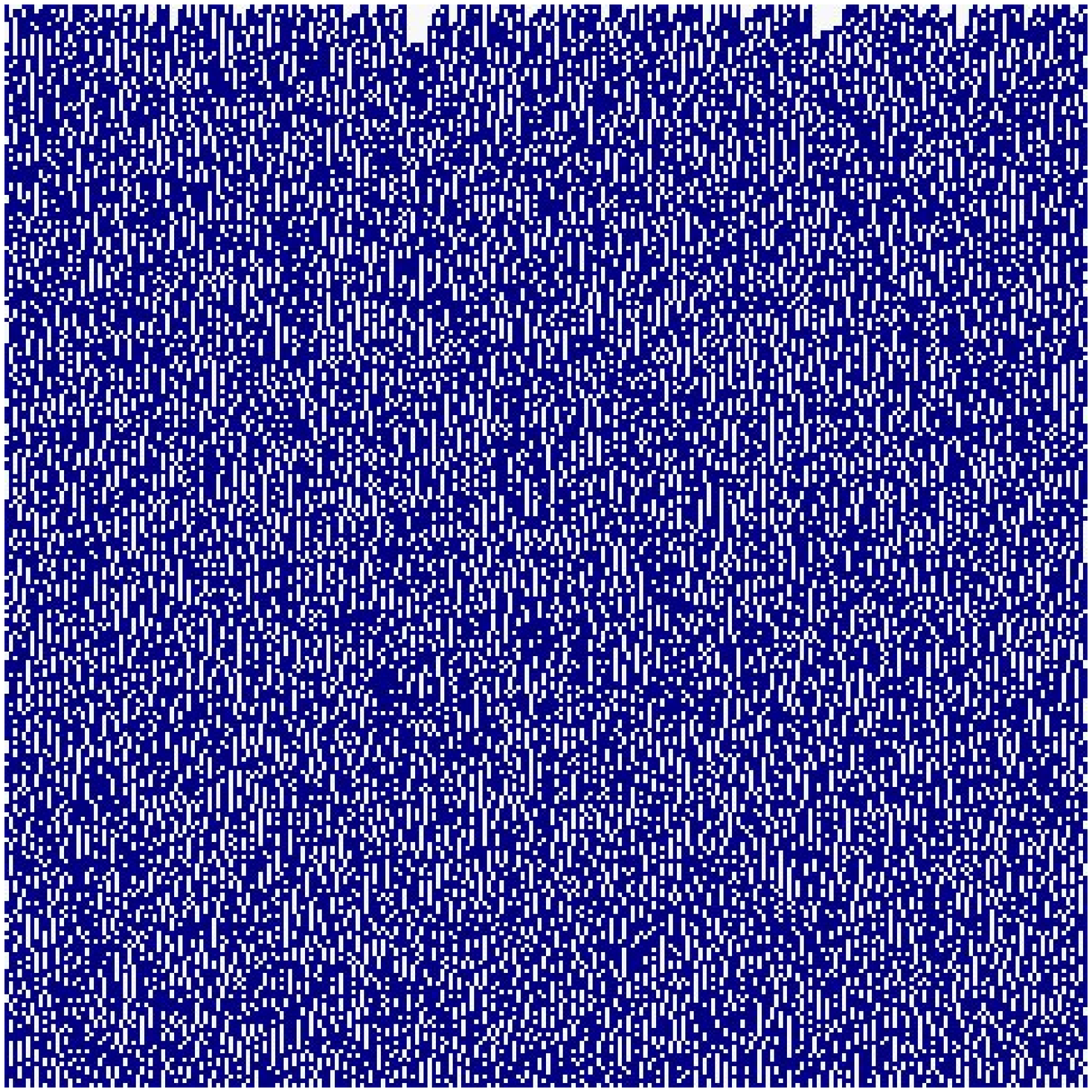}}
  \subfigure[unfair err]{
    \label{fig:110:d}
    \includegraphics[width=.21\textwidth]{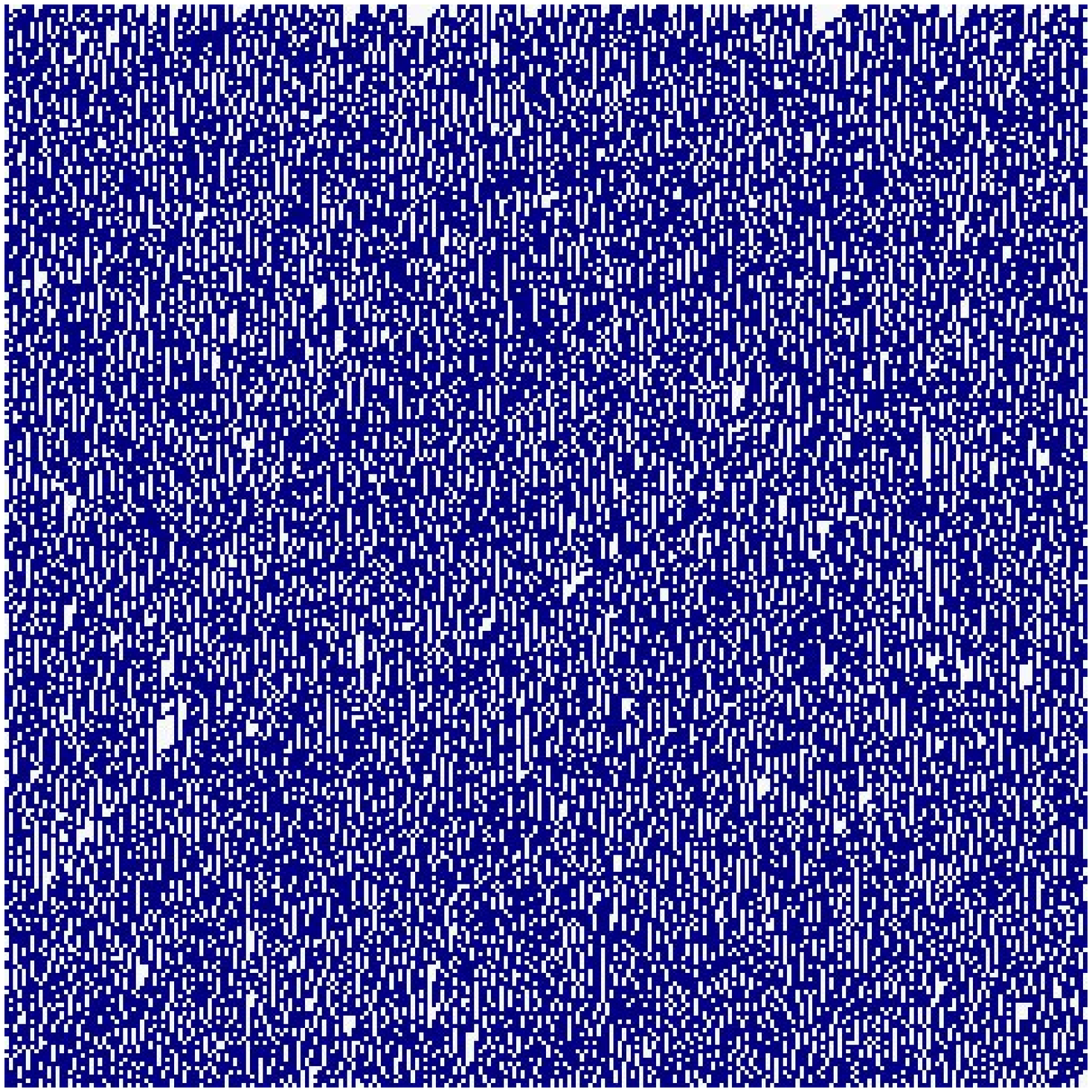}}
  \caption{Rule 110.  The error is asymmetric, $p_{10}=1\%$ and
    $p_{01}=0$.  CA with 256 cells.  Time grows down for 256
    iterations.}
  \label{fig:110}
\end{figure}

The synchronous updating is much sensitive to noise in the local
function, which is specially visible in complex and in periodic rules.
For chaotic and fixed point rules its behaviour is more stable.

In Fig. \ref{fig:110} we show results for rule 110, complex.  It can
be clearly noticed that the complex character of the rule in
synchronous update changed into a chaotic behaviour with only an error
of 1\% (1$\rightarrow$0).  The behaviour of the asynchronous unfair
update is approximately the same in the two cases.

\begin{figure}[center]
  \centering
  \subfigure[sync]{
    \label{fig:38:a}
    \includegraphics[width=.21\textwidth]{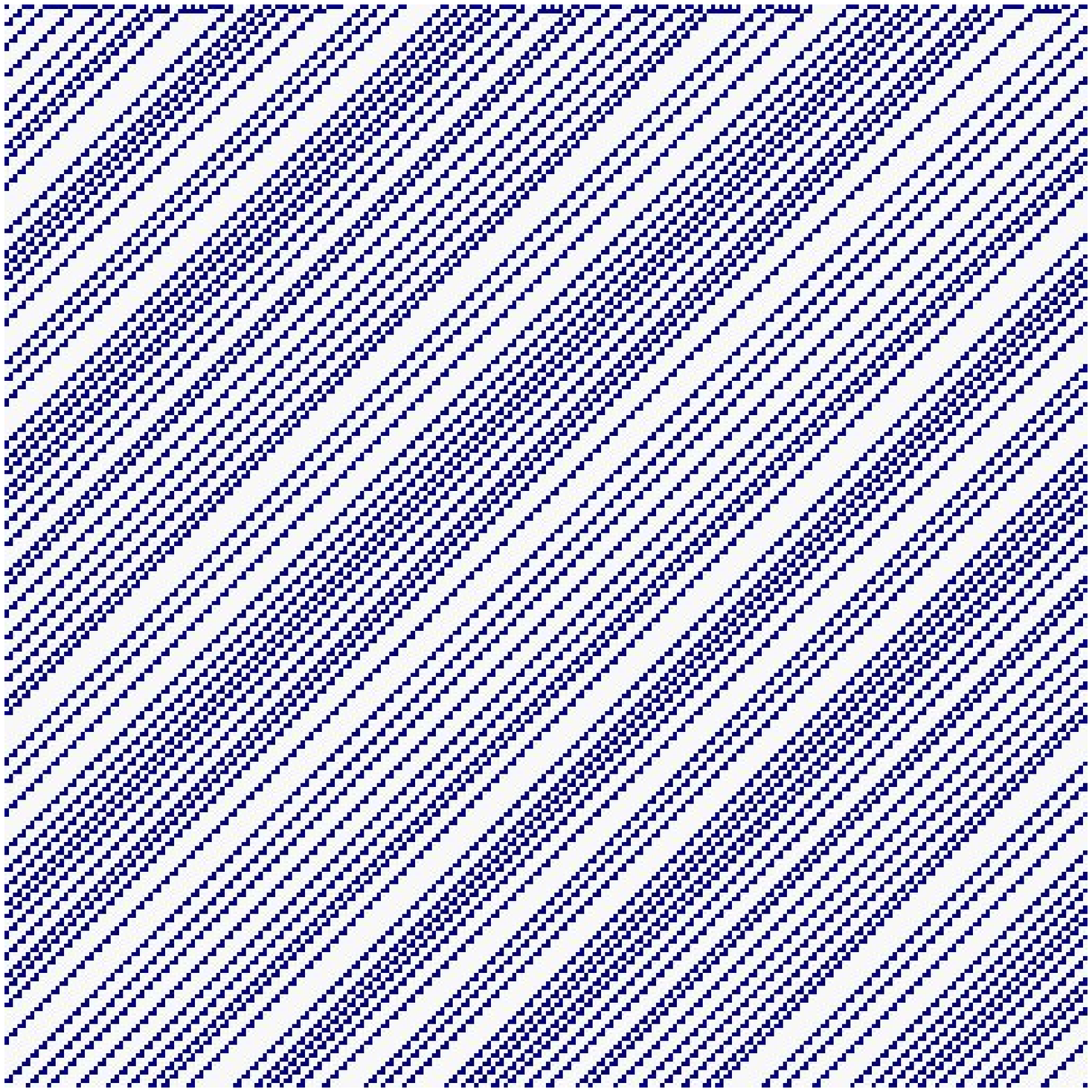}}
  \subfigure[sync err]{
    \label{fig:38:b}
    \includegraphics[width=.21\textwidth]{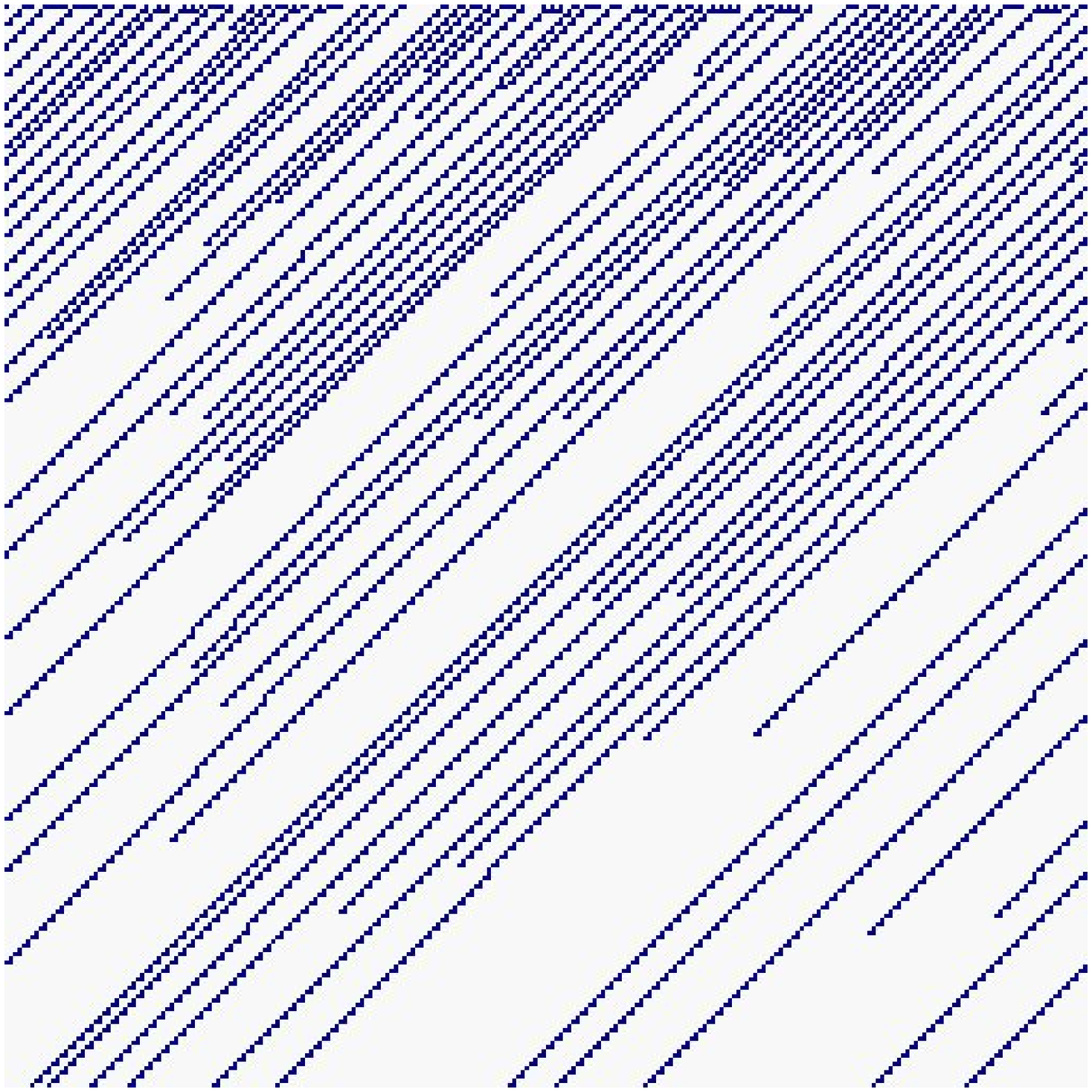}}
  \hspace{0.3cm}
  \subfigure[unfair]{
    \label{fig:38:c}
    \includegraphics[width=.21\textwidth]{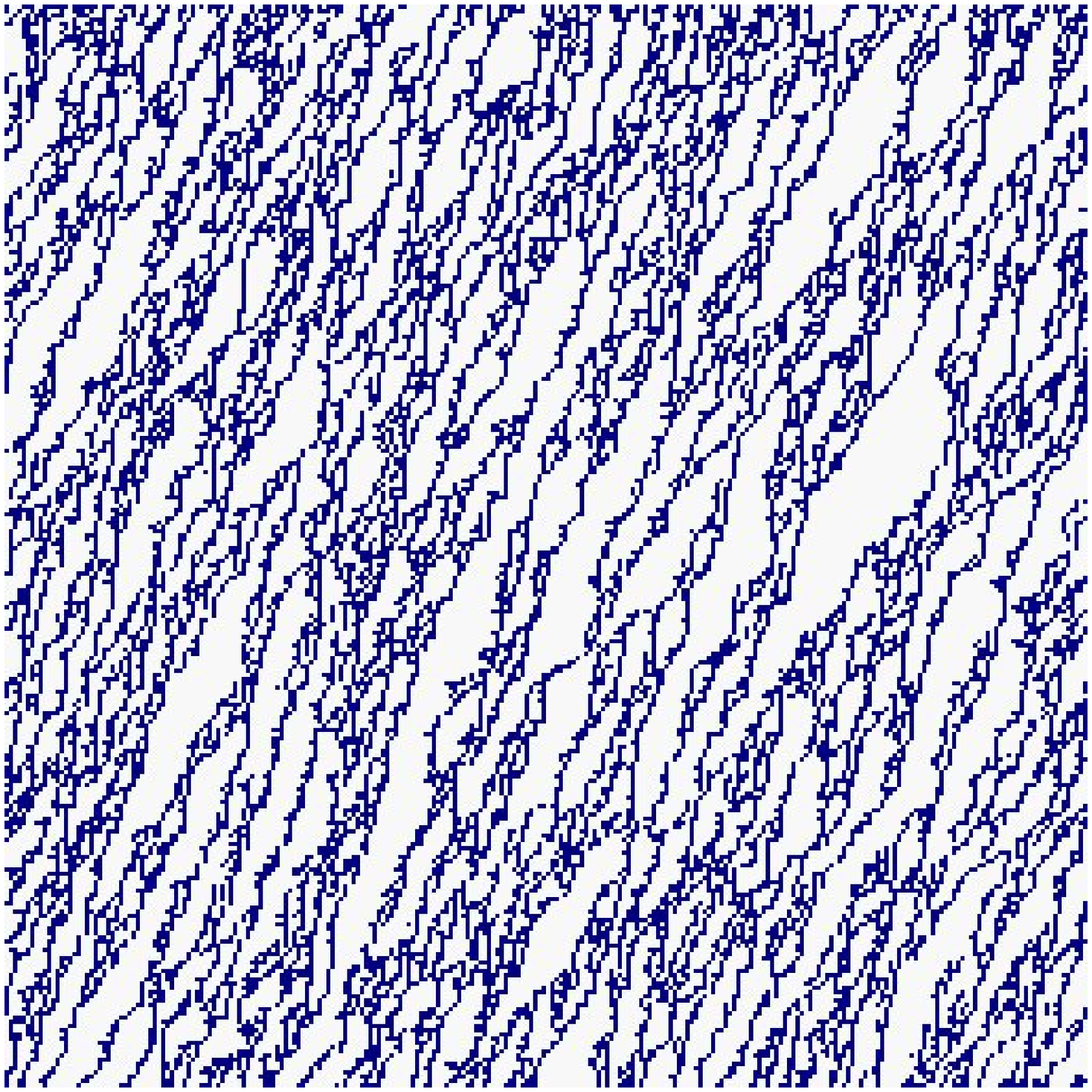}}
  \subfigure[unfair err]{
    \label{fig:38:d}
    \includegraphics[width=.21\textwidth]{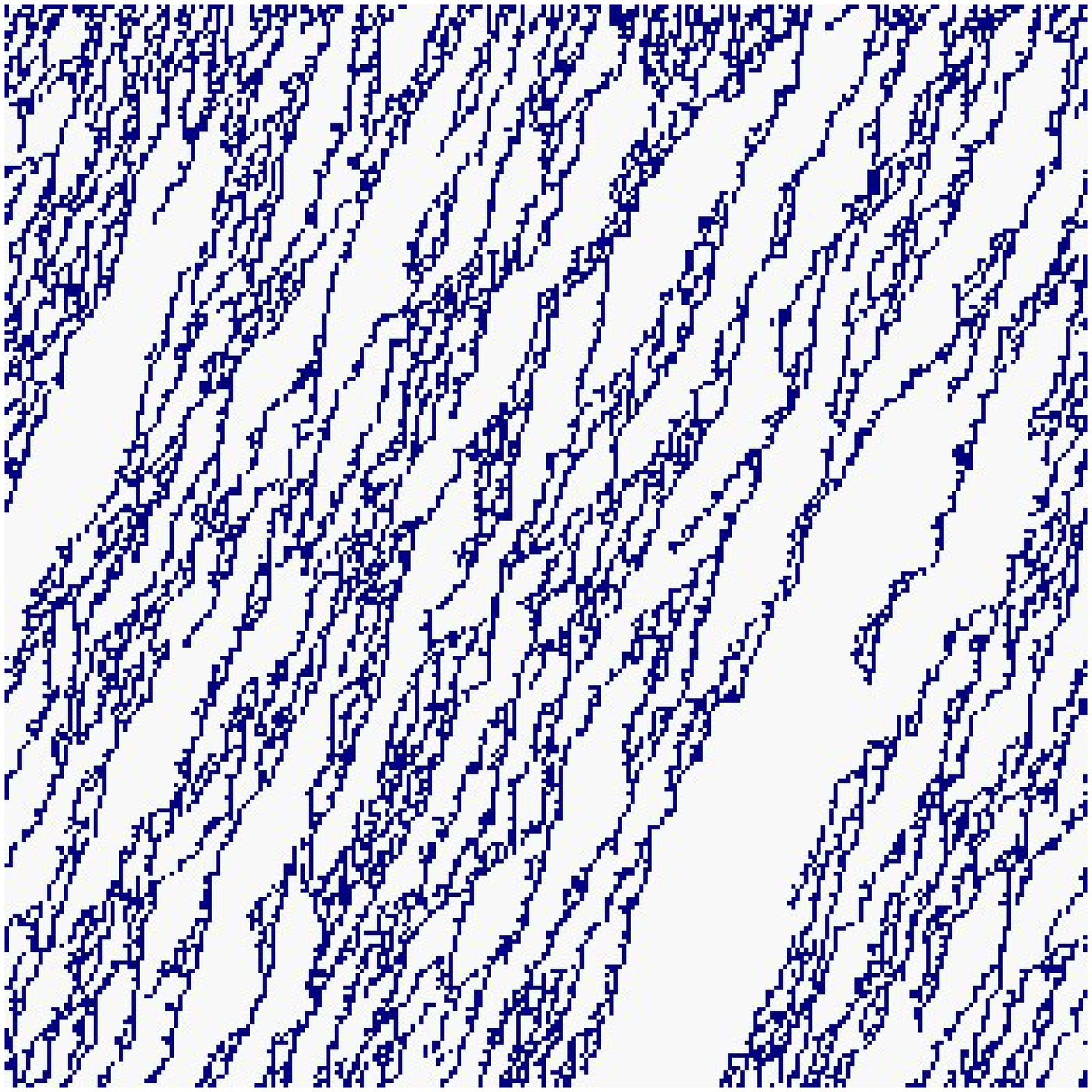}}
  \caption{Rule 38.  The error is asymmetric, $p_{10}=1\%$ and
    $p_{01}=0$.  CA with 256 cells.  Time grows down for 256
    iterations.}
  \label{fig:38}
\end{figure}

The periodic rule 38 is depicted in Fig. \ref{fig:38}.  The small
amount of noise (1\%) in the local function is enough to destroy the
periodic behaviour of the synchronous update.  In fact, after some
more iterations it converges to an all zero state.  Again, the
behaviour of the asynchronous unfair update does not sensibly change
with the addition of errors in the local function output.

Notice, in both cases, that the behaviour of the two updatings is
quite different already in the case with no errors.  Asynchronous
updates, in general, present much less structure in the patterns.

\section{Discussion}

It was already known that synchronous updates artificially introduce
structure in the patterns displayed by CA.  In this work we show that
the introduction of errors in the local function is also critical for
the structure of synchronous updates.

Errors in the local function, however, result in qualitatively
different behaviour from asynchronous updates.  A synchronous update
with local function errors still maintains some amount of structure
correlated with the errorless synchronous update.  This effect is
quite resilient.  To eliminate it we need to increase error percentage
to very high values, near random noise.

A negative answer can thus be given to the hypothesis raised in
\cite{sch_roo:99} of stochasticy acting the same way unregarding
whether it is introduced via asynchronous update or via stochastic
local functions.  The two processes are qualitatively different, in
spite of both being able to modify the behaviour of the artificial
synchronous updating model.  Synchronous CA are very sensitive to
noise, whether in amplitude (local function) or in time (asynchrony),
but in different ways.

\bibliography{noise-ca}
\bibliographystyle{splncs}

\end{document}